\documentclass[prc,aps,epsfig]{revtex4}
\usepackage{epsfig}
\def\beq{\begin{equation}}
\def\enq{\end{equation}}
\def\beqa{\begin{eqnarray}}
\def\enqa{\end{eqnarray}}
\def\MeV{\nobreak\,\mbox{MeV}}
\def\GeV{\nobreak\,\mbox{GeV}}

\def\pli{p^\prime}

\def\la{\lambda}

\def\Ga{\Gamma}

\def\qq{\lag\bar{q}q\rag}

\def\G3{\lag g^3G^3\rag}
\def\lb{\label}

\newcommand{\rag}{\rangle}
\newcommand{\lag}{\langle}

\begin{document}
\title{Investigating the tetraquark structure of the new mesons}
\author{M. Nielsen$^1$, F.S. Navarra$^1$ and M.E. Bracco$^2$}
\affiliation{$^1$Instituto de F\'{\i}sica, 
        Universidade de S\~{a}o Paulo,
  C.P. 66318,  05389-970 S\~{a}o Paulo, SP, Brazil\\
$^2$Instituto de F\'{\i}sica, Universidade do Estado do Rio de 
Janeiro, 
Rua S\~ao Francisco Xavier 524, 20550-900 Rio de Janeiro, RJ, Brazil}
      
\begin{abstract}
Using the QCD sum rule approach we investigate the 
possible four-quark structure of the recently observed mesons 
$D_{sJ}^{+}(2317)$, firstly observed by BaBaR, $X(3872)$, firstly
observed by BELLE and $D_0^{*0}(2308)$ observed by BELLE. We use  
diquark-antidiquark currents and work  in full QCD, without relying on 
$1/m_c$ expansion. Our results indicate that a four-quark structure is 
acceptable for these mesons.
\end{abstract}

\maketitle
The recent observations of the very narrow resonances
$D_{sJ}^+(2317)$ by BaBar \cite{babar},
$D_{sJ}^{+}(2460)$ by CLEO \cite{cleo}, and $X(3872)$ by BELLE \cite{BELLE},
all of them with masses below quark model predictions, have stimulated a
renewed interest in the spectroscopy of open charm and charmonium states.
Due to their narrowness and small masses, these new mesons were 
considered as good candidates for  four-quark states  by many authors
\cite{Swanson}. The idea of mesons as four-quark states is not new. Indeed, 
even Gell-Mann
in his first work about quarks had mentioned that mesons could
be made out of $(q\bar{q}),~(qq\bar{q}\bar{q})$ {\it etc.} \cite{gell}. The 
best known example of applying the idea of four-quark states for mesons is 
for the light scalar mesons (the isoscalars $\sigma(500),~f_0(980)$, the 
isodublet $\kappa(800)$ and the isovector $a_0(980)$) \cite{jaffe,cloto}. 
In a four-quark scenario, the mass degeneracy of $f_0(980)$ and $a_0(980)$ is 
natural, the mass hierarchy pattern of the nonet is understandable, and
it is easy to explain why $\sigma$ and $\kappa$ are broader than $f_0(980)$ 
and $a_0(980)$. 

In refs.~\cite{blmnn,x3872} the  method of QCD  sum rules (QCDSR) 
\cite{svz,rry,narison} was used to study the two-point functions for the mesons
$D_{sJ}^+(2317)$ and $X(3872)$ considering them as four-quark states 
in a  diquark-antidiquark configuration.
The results obtained for their masses are given in Table I. 
\begin{center}
\small{{\bf Table I:} Numerical results for the resonance masses}
\\
\vskip3mm

\begin{tabular}{|c|c|c|}  \hline
resonance & $D_{sJ}$ & $X$   \\
\hline
mass (GeV) &$2.32\pm0.13$  & $3.93\pm0.15$\\
\hline
\end{tabular}\end{center}

Comparing the results in Table I with the resonance masses given 
by: $D_{sJ}^+(2317)$ and $X(3872)$, we see that it is possible to reproduce 
the experimental value of the masses using a four-quark representation for 
these states. 

The study of the three-point functions related to the
decay widths  $D_{sJ}^+(2317)\to D_s^+\pi^0$ and 
$X(3872)\to J\psi\pi^+\pi^-$,
using the  diquark-antidiquark configuration for $D_{sJ}$ and $X$, was 
done in refs.~\cite{decayds,decayx}.  The results obtained for their partial
decay widths are given in Table II. 
\begin{center}
\small{{\bf Table II:} Numerical results for the resonance decay widths}
\\
\vskip3mm

\begin{tabular}{|c|c|c|}  \hline
decay & $D_{sJ}^+\to D_s^+\pi^0$ & $X\to J/\psi\pi^+\pi^-$   \\
\hline
$\Gamma$ (MeV) &$(6\pm2)\times10^{-3}$  & $50\pm15$\\
\hline
$\Gamma^{exp}_{tot}$ (MeV) &$<5$  & $<2.3$\\
\hline
\end{tabular}\end{center}

From Table II we see that the partial decay width obtained in 
ref.~\cite{decayds}, supposing that the mesons  $D_{sJ}^+(2317)$ is a 
four-quark state, is consistent with the experimental upper limit. However,
in the case of the meson $X(3872)$, the  partial decay width obtained in 
ref.~\cite{decayx} is much bigger than the experimental upper limit to the
 total width. 

In ref.~\cite{decayx} some arguments were presented to reduce the value
of this decay width, by imposing that the initial four-quark state
needs to have a non-trivial color structure. In this case, its partial decay 
width can be reduced to
$\Ga(X\to J/\psi\pi^+\pi^-))=(0.7\pm0.2)~\MeV$. However,
that procedure may appear somewhat unjustified and, therefore, more study
is needed until one can arrive at a definitive conclusion about the 
structure of the meson $X(3872)$.

In ref.~\cite{blmnn}, besides the four-quark state $(cq)(\bar{s}\bar{q})$
representing the meson $D_{sJ}^+(2317)$, it was also studied the configuration
$(cq)(\bar{u}\bar{d})$ associated with a possible scalar meson that we
will call $D(0^+)$ (the  $0^+$ stands for $J^P$). The mass obtained for this 
state is: $m_{D(0^+)}=(2.22\pm0.21)$ MeV, in a very good agreement with the
prediction made in ref.~\cite{bardeen} for the $D(0^+)$ scalar meson:
$m_{D(0^+)}=(2.215\pm0.002)$ MeV. This value was obtained in 
ref.~\cite{bardeen} by supposing that the meson $D(0^+)$ is the chiral 
partner of the meson $D$, with the same mass difference as the chiral 
pair $D_{sJ}^+(2317)-D_s$. The authors of ref.~\cite{bardeen} have also 
evaluated the decay widths $D_{sJ}^+\to D_s^+\pi^0$ and $D(0^+)\to D\pi^\pm$
obtaining: $\Gamma(D_{sJ}^+\to D_s^+\pi^0)=21.5G_A^2$ keV and
$\Gamma(D(0^+)\to D\pi^\pm)=326G_A^2$ MeV, where they expect $G_A\sim1$.

Here, we extend the calculation done in refs.~\cite{blmnn,decayds} to study 
the vertex associated with the decay $D^0(0^+)\to D^+\pi^-$.
The QCDSR calculation for the vertex, $D^0(0^+) D^+\pi^-$, centers
around the three-point function given by
\beq
T_\mu(p,\pli,q)=\int d^4x d^4y ~e^{i\pli.x}~e^{iq.y}
\lag 0 |T[j_{D}(x)j_{5\mu}(y)j^\dagger_0(0)]|0\rag,
\lb{3po}
\enq
where $j_0$ is the interpolating field for the scalar $D^0(0^+)$ meson
\cite{blmnn}:
\beq
j_0=\epsilon_{abc}\epsilon_{dec}\left[(d_a^TC\gamma_5c_b)
(\bar{u}_d\gamma_5C\bar{d}_e^T)\right],
\label{int}
\enq
where $a,~b,~c,~...$ are colour indices and $C$ is the charge conjugation
matrix. 
In Eq.~(\ref{3po}), $p=\pli+q$ and the interpolating fields for the $\pi^-$
and $D^+$ mesons are given by:
\beq
j_{5\mu}=\bar{u}_a\gamma_\mu\gamma_5d_a
,\,\;\;\;j_{D}=i\bar{d}_a\gamma_5c_a.
\lb{pseu}
\enq

The calculation of the phenomenological side proceeds by 
inserting intermediate states for $D$, $\pi$ and $D(0^+)$, and by using 
the definitions: 
$\lag 0 | j_{5\mu}|\pi(q)\rag =iq_\mu F_{\pi}$,
$\lag 0 | j_{D}|D(\pli)\rag ={m_{D}^2f_{D}\over m_c}$,
$\lag 0 | j_{0}|D(0^+)(p)\rag =\la_0$.
We obtain the following relation:
\beq
T_{\mu}^{phen} (p,\pli,q)={\la_0 m_{D}^2f_{D}F_{\pi}~
g_{D(0^+)D\pi}/m_c
\over (p^2-m_{D(0^+)}^2)({\pli}^2-m_{D}^2)(q^2-m_\pi^2)}~
~q_\mu +\mbox{continuum contribution}\;,
\lb{phen}
\enq
where the coupling constant, $g_{D(0^+)D\pi}$, is defined by the 
on-mass-shell matrix element: $\lag D \pi|D(0^+)\rag=g_{D(0^+)D\pi}$.
The continuum contribution in Eq.(\ref{phen}) contains the contributions of
all possible excited states.

In the case of the light scalar mesons, considered as diquark-antidiquark 
states, the study of their vertices functions using the QCD sum rule approach 
at the pion pole \cite{narison,rry,nari2}, was done in ref.\cite{sca}. It was
shown that  the decay widths determined from the QCD sum rule calculation are 
consistent with existing experimental data. 
Here, we follow refs.~\cite{decayds,sca} and work at the pion pole. 
The main reason for working at the pion pole is that  one does not 
have to deal with the complications associated with the extrapolation of the 
form factor \cite{dosch}. The pion pole method consists in neglecting the pion 
mass in the denominator of Eq.~(\ref{phen}) and working at $q^2=0$. In the 
OPE side one singles out the leading terms in the operator product expansion 
of Eq.(\ref{3po}) that match the $1/q^2$ term. Since we are working at 
$q^2=0$, we take the limit $p^2={\pli}^2$ and we 
apply a single Borel transformation to $p^2,{\pli}^2\rightarrow M^2$.
In the phenomenological side, in the structure $q_\mu$ we get \cite{decayds}:
\beq
T^{phen}(M^2)= {\la_0 m_{D}^2f_{D}F_{\pi}~
g_{D(0^+)D\pi}\over m_c(m_{D(0^+)}^2-m_{D}^2)}\left(
e^{-m_{D}^2/M^2} -e^{-m_{D(0^+)}^2/M^2}\right)+A~e^{-s_0/M^2}+
\int_{u_0}^\infty\rho_{cc}(u)~e^{-u/M^2}du,
\label{paco}
\enq
where $A$ and $\rho_{cc}(u)$
stands for the pole-continuum transitions and pure continuum contributions,
with $s_0$ and $u_0$ being the continuum thresholds for $D(0^+)$ and $D$ 
respectively. For simplicity, one assumes that the pure continuum 
contribution to the spectral density, $\rho_{cc}(u)$, is given by the result 
obtained in the OPE side. 
Therefore, one uses the ansatz: $\rho_{cc}(u)=\rho_{OPE}(u)$.
In Eq.(\ref{paco}), $A$ is a parameter which, together with
$g_{D(0^+)D\pi}$, has to be determined by the sum rule.

In the OPE side we single out the leading terms proportional to
$q_\mu/q^2$.  Transferring the pure continuum contribution to
the OPE side, the sum rule for the coupling constant, up to dimension 7, is 
given by:
\beq
C~\left(e^{-m_{D}^2/M^2} -e^{-m_{D(0^+)}^2/M^2}\right)+A~e^{-s_0/M^2}=
2\qq\left[{1\over2^4\pi^2}\int_{m_c^2}^{u_0}du~e^{-u/M^2}u\left(1-
{m_c^2\over u}\right)^2-{m_c\qq\over6}e^{-m_c^2/M^2}\right],
\label{sr}
\enq
with 
\beq
C={\la_0 m_{D}^2f_{D}F_{\pi}
\over m_c(m_{D(0^+)}^2-m_{D}^2)}~g_{D(0^+)D\pi}.
\label{coef}
\enq

In the numerical analysis of the sum rules, the values used for the meson
masses, quark masses and condensates are: $m_{D(0^+)}=2.2~\GeV$,
$m_{D}=1.87~\GeV$, $m_c=1.2\,\GeV$, 
$\lag\bar{q}q\rag=\,-(0.23)^3\,\GeV^3$. For the meson
decay constants we use $F_\pi=\sqrt{2}~93\MeV$ and $f_{D}=0.20~\GeV$
\cite{cleo2}. We use $u_0=6~\GeV^2$ and
for the current meson coupling, $\la_0$, we are going
 to use the result obtained from the two-point function in ref.~\cite{blmnn}.
Considering $2.6\leq\sqrt{s_0}\leq2.8~\GeV$ we get $\la_0=(3.3\pm0.3)\times
10^{-3}~\GeV^5$.
\begin{figure}[h]
\centerline{\epsfig{figure=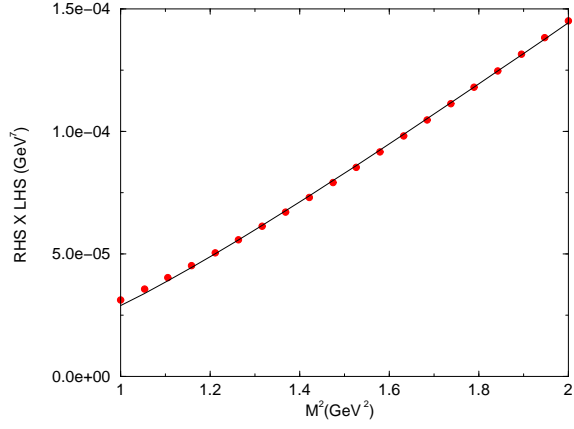,width=7.5cm}}
\caption{\small{{Dots: the RHS of Eq.(\ref{sr}), as a function of the Borel 
mass.  The solid line gives the fit of the QCDSR results through 
the LHS of Eq.(\ref{sr}).}}}
\label{fig1}
\end{figure}

In Fig.~1 we show, through the dots, the right-hand side (RHS) of 
Eq.(\ref{sr}) as a function of the Borel mass. We use the same Borel window 
as defined in ref.\cite{blmnn}.
To determine $g_{D(0^+)D\pi}$ we fit the QCDSR results with the analytical
expression in the left-hand side (LHS) of Eq.(\ref{sr}).
Using $\sqrt{s_0}=2.7\GeV$ we get: $C=1.25\times10^{-3}~\GeV^7$ and 
$A=1.47\times10^{-3}~\GeV^7$. Using the definition of $C$ in Eq.(\ref{coef})
and $\la_0=3.3\times10^{-3}~\GeV^5$ (the value obtained for $\sqrt{s_0}=2.7
\GeV$) we get $g_{D(0^+)D\pi}=6.94~\GeV$. Allowing $s_0$ to vary in the
interval $2.6\leq\sqrt{s_0}\leq2.8~\GeV$, the corresponding variation
obtained for the coupling constant is $5~\GeV\leq g_{D(0^+)D\pi}\leq
7.5~\GeV$.

The coupling constant, $g_{D(0^+)D\pi}$, 
is related to the partial decay width through the relation:
\beq
\Gamma(D^0(0^+)\rightarrow D^+\pi^-)={1\over 16\pi m_{D(0^+)}^3}
g_{D(0^+)D\pi}^2\sqrt{\la(m_{D(0^+)}^2,m_{D}^2,m_{\pi}^2)},
\lb{decay}
\enq
where $\la(a,b,c)=a^2+b^2+c^2-2ab-2ac-2bc$. Allowing $s_0$ to vary in the 
range discussed above we get:
\beq
\Gamma(D^0(0^+)\rightarrow D^+\pi^-)=(120\pm20)\MeV.
\label{fin}
\enq

In Table III we show the partial decay width obtained in ref.~\cite{bardeen},
in ref.~\cite{decayds} and here for different decays.
From the results in Table III we see that if one uses $G_A=0.6$, the result
presented here and the result in ref.~\cite{decayds} are consistent with
the results presented in ref.~\cite{bardeen} for both decays.

\begin{center}
\small{{\bf Table III:} Numerical results for the resonance partial decay 
widths from different approaches}
\\
\vskip3mm

\begin{tabular}{|c|c|c|c|}  \hline
decay & ref.~\cite{bardeen} & ref.~\cite{decayds} & this work  \\
\hline
$D_{sJ}^+\to D_s^+\pi^0$ &$21.5~G_A^2$ keV  & $(6\pm2)$ keV 
&\\
\hline
$D^0(0^+)\to D^+\pi^-$ &$326~G_A^2$ MeV  &  &$120\pm20$ MeV\\
\hline
\end{tabular}\end{center}

It is important to notice that the
BELLE Collaboration \cite{belle2} has  reported the observation of 
a rather broad scalar meson $D_0^{*0}(2308)$ in the decay mode $D_0^{*0}(2308)
\to D^+\pi^-$ with a total width $\Gamma\sim270~\MeV$. Although both, 
the mass and the total decay width reported in \cite{belle2}, are bigger than 
the values found for the meson $D(0^+)$ studied here, we can not discard the 
possibility that the BELLE's resonance can be interpreted as a four-quark 
state.

We have presented a QCD sum rule study of the vertex function associated with
the strong decay $D^0(0^+)\rightarrow D^+\pi^-$, where the scalar
$D(0^+)$ meson was considered as diquark-antidiquark state. We get for the 
partial decay width: $\Gamma(D^0(0^+)\rightarrow D^+\pi^-)=(120\pm20)~\MeV$.

\vspace{1cm}
 
\underline{Acknowledgements}: 
This work has been supported by CNPq and FAPESP. 

\end{document}